\documentclass[twocolumn]{article}

\usepackage{lipsum}
\usepackage{hyperref,hhline,amssymb,colortbl,cals,multirow,graphicx,amsmath,nccmath}
\usepackage{tabularx,float,textcomp}
\usepackage{algorithm}
\usepackage[square,sort,comma,numbers]{natbib}
\usepackage[noend]{algpseudocode}

\newcommand\Tstrut{\rule{0pt}{2.6ex}}
\newcommand\Bstrut{\rule[-0.9ex]{0pt}{0pt}}

\title{Generating Realistic Two-Line Elements for Notional Space Vehicles and Constellations}
\author{Troy Rockwood, Greg Steeger, Matthew Stein}

\begin{document}


\twocolumn[
\begin{@twocolumnfalse}
	\maketitle
	\begin{abstract}
		\normalsize As space becomes increasingly populated with new satellites and systems, modeling and simulating existing and future systems becomes more important.  The two-line element set has been a standard format for sharing data about a satellite's orbit since the 1960s, and well-developed algorithms can predict the future location of satellites based on this data.  
		In order to simulate potential future systems, especially when mixed with existing systems, data must be generated to represent the desired orbits.  
		We present a means to create two-line element sets with parameters that closely resemble real satellite behavior, and rely on a novel approach to calculate the mean motion for even greater accuracy.
		\vspace{20px}
	\end{abstract}
\end{@twocolumnfalse}
]

\section{Introduction}

The population of active satellites in orbit around Earth has increased exponentially over the last few decades\cite{UCSSatelliteDatabase_2021}. Commercial, civil, and government entities continue to expand their role and influence in space.  Modeling and simulation of satellites and their behavior has likewise increased.  The use cases include constellation planning, optimization, performance assessment, and conjunction analysis\cite{Coverage1,8396743,Design}.  

There are many software tools and algorithms to simulate satellites and constellations of satellites. Among these is the well-known simplified perturbations model, often referred to as SGP4\cite{SGP4}.  These algorithms make use of two-line element sets (TLEs)\cite{kelso_2019,ValladoTLE} to describe the orbit of each satellite and begin the simulation.  Regardless of the algorithm employed, TLEs are widely used and published for most active and inactive objects in orbit.

For the purposes of planning future space systems, it is useful to simulate satellites and orbits that do not yet exist, possibly alongside real systems already in orbit. 
Doing so requires the generation of TLEs for the notional systems.  
A naïve approach based only on classical orbital elements might look like the following:
\begin{enumerate}
	\item The desired inclination, eccentricity, altitude at perigee, and argument of perigee are chosen according to the desired orbit
	\item The first and second derivative of mean motion and drag terms are set to exactly zero
	\item The mean motion is calculated from Kepler's third law using the altitude at perigee and a mean radius of the Earth of 6,371\,km
	\item The rest of the TLE elements are freely chosen, and the checksum is computed for the last character in each line
\end{enumerate}

The classical orbital elements in Step\,1 are freely chosen, but Step\,2 and\,3 each present opportunities for significant error in the desired simulation results.  When propagating TLEs, some or all of the terms in Step\,2 help model realistic orbital decays depending on the algorithm chosen.  Simulated notional TLEs may thus behave quite differently than TLEs created for orbiting satellites, making mixed simulations less realistic especially as simulation time increases.  Similarly, calculating mean motion from a fixed value for the radius of the Earth presents potential for significant error in drift.  Because the Earth is not a perfect sphere, gravitational effects experienced by satellites will be a function of the intended orbit, namely the inclination, eccentricity, and argument of perigee.

In order to realistically simulate the behavior of satellites, the generated TLEs should contain terms that accurately resemble those of real systems.
We describe a novel method for generating TLEs for almost any notional system including realistic terms for drag and the first and second derivative of mean motion.  Additionally, we describe a novel way to determine the mean motion (MM) for the desired orbit which is a function of inclination, eccentricity, and the argument of perigee.  Finally, we summarize our approach which is an expansion of the above steps in Section~\ref{sec:assemble}, including a polynomial approach to determining the mean radius of the Earth for computing the mean motion. 

\section{Essential TLE Terms}

This study ignores the title line sometimes included in some TLEs, and focuses only on lines one and two. A detailed visualization and explanation of each TLE term is illustrated in Figure~1 of Ref~\cite{kelso_2019}. The satellite catalog number, classification, international designator value, element set number, and revolution number can all be chosen at will, and do not affect the simulation.  The only exception is that satellite catalog numbers should not conflict with other satellites in the simulation. Additionally, the ephemeris type should be 0.

The epoch of the TLE can be chosen at will, but in generating an epoch one should consider the following implications.  If a simulation only involves notional systems, then it is only important that all TLEs in the simulation have a similar epoch.
If mixing notional with real satellites, it is best to choose an epoch close to the real systems of interest.

The inclination, right ascension of the ascending node (RAAN), eccentricity, argument of perigee, and mean anomaly are all chosen according to the desired simulation.  For example, a notional low Earth orbit constellation might consist of 225 satellites in 15 planes with 15 satellites per plane.  The inclination might be chosen to be {90.0\textdegree} to model a polar orbit.  To provide somewhat even distribution of coverage around the equator, the RAAN of each plane could range from {0.0\textdegree} to {168.0\textdegree} in steps of {12.0\textdegree}. Within the planes, the mean anomaly might likewise be spaced from {0.0\textdegree} to {336.0\textdegree} in steps of {24.0\textdegree}.  

\section{Catalog-Derived Terms}

Three of the remaining terms--the first and second derivative of the mean motion, and the drag term--can be derived from existing TLEs which are accessible in open repositories on the internet.  We use a collection of 18,477 published TLEs associated with satellites, rocket bodies, and other debris.  In order to assign terms to notional TLEs that represent realistic TLEs, the TLEs are categorized into low-Earth orbit (LEO), medium Earth orbit (MEO), geosynchronous (GEO), and highly elliptical orbit (HEO) populations.  

The algorithm to assign TLEs to a population is given in Algorithm~\ref{alg:TLEClassification} and the impact of the included criteria is illustrated in Figure~\ref{im_Orbit_Ecc_ListPlot}.  There are various other criteria and algorithms, and the methods in this paper could be easily adapted by modifying Algorithm~\ref{alg:TLEClassification}, including additional population categories.

\begin{figure}[h]
	\begin{algorithm}[H]
		\caption{TLE Classification}
		\label{alg:TLEClassification}
		\begin{algorithmic}
			\If{Eccentricity {$\geq$} 0.5}
				\State HEO			
			\ElsIf{Mean Motion {$\geq$} 11.25}
				\State LEO
			\ElsIf{Mean Motion {$\geq$} 1.2}
				\State MEO
			\Else
				\State GEO
			\EndIf
		\end{algorithmic}
	\end{algorithm}
\end{figure}

\begin{figure}
	\caption{The categorization of 18,477 published TLEs using the criteria from Algorithm~\ref{alg:TLEClassification}.}
	\includegraphics[width=\linewidth]{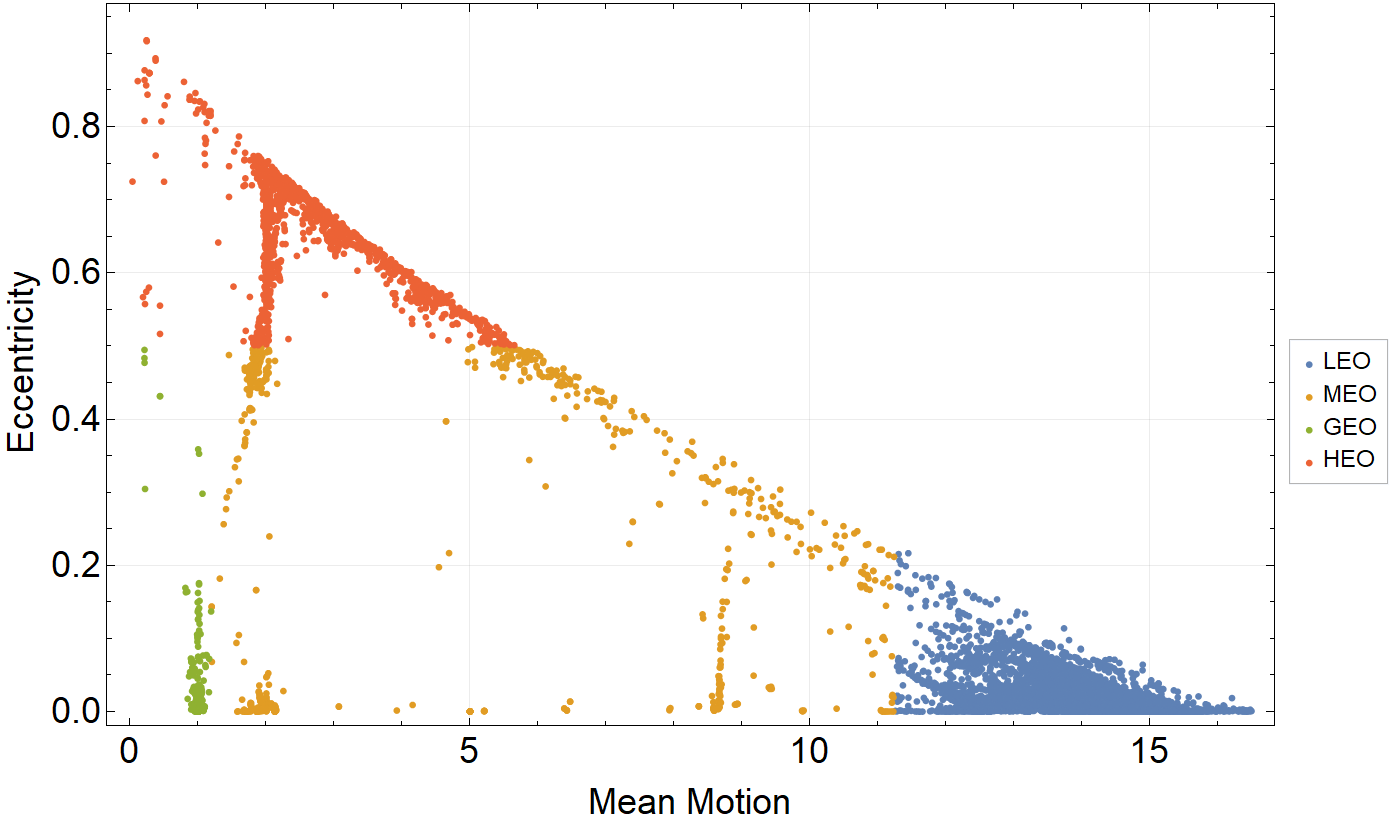}
	\label{im_Orbit_Ecc_ListPlot}
\end{figure}

\begin{figure}[!tb]
	\caption{The first and second derivative of the mean motion, and the drag term, highlighted by population type for all 18,477 TLEs in this study.}
	\includegraphics[width=\linewidth]{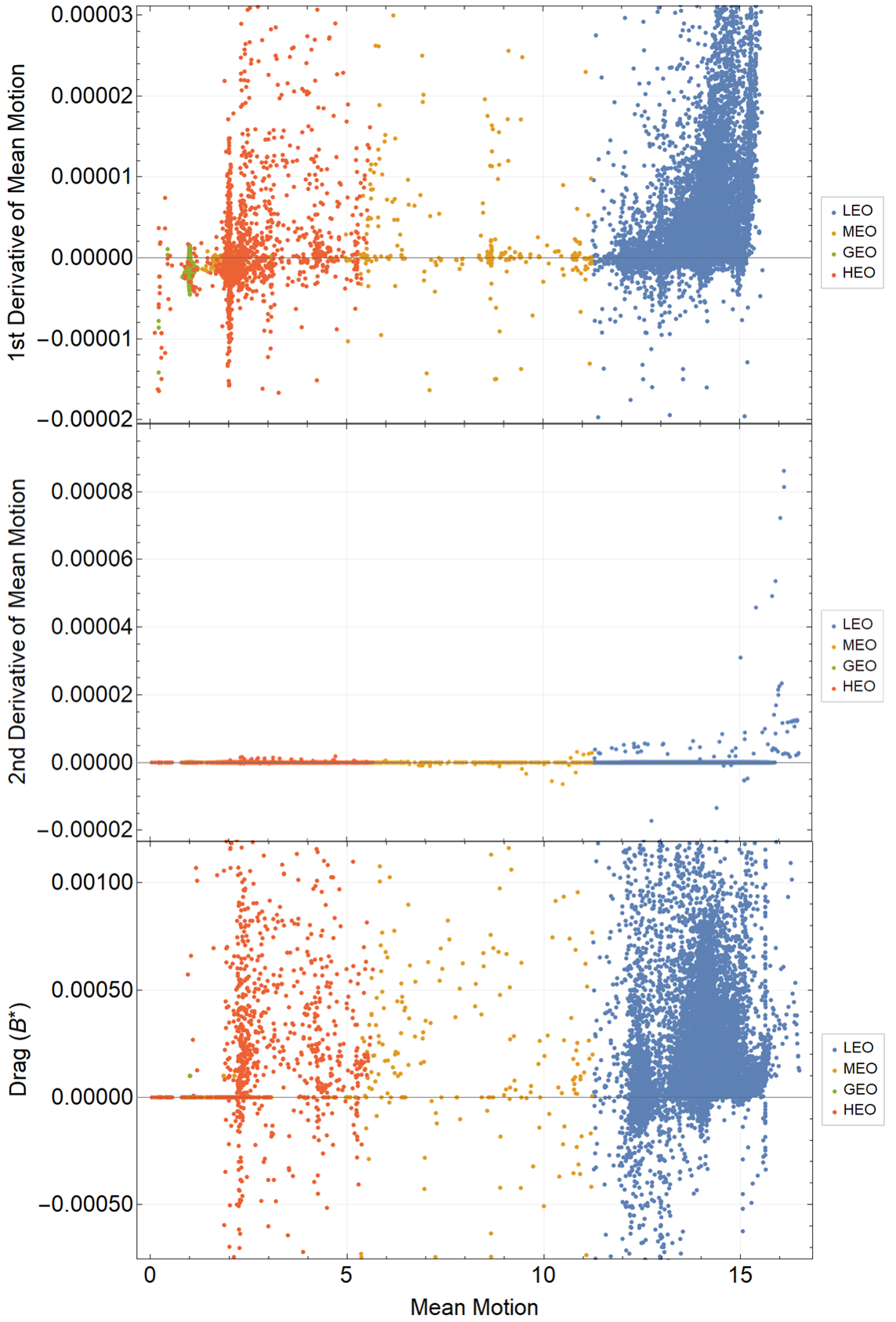}
	\label{im_Orbit_MM12B}
\end{figure}

With populations assigned, a mean value can be computed for the first and second derivative of the mean motion, and the drag term as shown in Table~\ref{table:1st2ndBstar} and Figure~\ref{im_Orbit_MM12B}.

\begin{table*}
	\centering
	\begin{tabular}{lccc}\hline
		\Tstrut Population & MM$'\,\times10^{3}~(\textrm{Rev}/\textrm{day}^2)$ & MM$''\,\times10^{6}~(\textrm{Rev}/\textrm{day}^3)$ & Drag Term$\,\times10^{3}~\left({R}_{\bigoplus}^{-1}\right)$\\
		\hline\Tstrut
		HEO & $0.048575\pm0.384281$ & $~~0.0125888\pm0.120716$ & $1.558450\pm27.4125$\\[1.9pt]
		LEO & $0.154256\pm3.778700$ & $~~0.0942242\pm2.866240$ & $0.377655\pm10.7492$\\[1.9pt]
		MEO & $0.154986\pm1.640110$ & $-0.0166109\pm0.381907$ & $1.295840\pm48.6634$\\[1.9pt]
		GEO & $0.001190\pm0.001467$ & $~~~~0\pm0$ & $0.639138\pm25.2697$\\\hline
	\end{tabular}
	\caption{The mean and standard deviation values for the first and second derivative of the mean motion (MM$'$ and MM$''$, respectively), and the drag term, grouped by population.\label{table:1st2ndBstar}}
\end{table*}

\section{Determining Mean Motion}

A simple modification of Kepler's third law allows the calculation of the mean motion in revolutions per day from the semi-major axis of orbit, as shown in Equation~\ref{eq:MM}:

\begin{equation}
	\label{eq:MM}
	\textrm{MM}=\frac{T_{\bigoplus}}{2\pi}\sqrt{\frac{\mu}{a^3}}
\end{equation}
where $T_{\bigoplus}$ is a solar day (86,400 seconds), $a$ is the semi-major axis, and $\mu$ is the standard gravitational parameter of Earth ($3.986004418\times10^{14} \textrm{m}^3/\textrm{s}^2$). The semi-major axis can be computed from the radius of the Earth $R_{\bigoplus}$, the altitude at perigee $A_p$, and the eccentricity $e$ as shown in Equation~\ref{eq:sma}.

\begin{equation}
	\label{eq:sma}
	a=\frac{A_p+R_{\bigoplus}}{1-e}
\end{equation}


The Earth's radius $R_{\bigoplus}$ is the distance from the center of the Earth to a point on its surface. Because the Earth is not a perfect sphere and due to the Earth's topography, this radius is not constant across the Earth's surface. The Earth can be modeled as an oblate spheroid with a minimum radius of 6,357\,km at the poles, and a maximum radius of 6,378\,km at the equator. For reference, analysis commonly relies on a nominal or ``globally averaged'' value  of 6,371\,km\citep{Mamajek2015,Moritz1980}. Globally averaged values are computed by either
\begin{itemize}
	\item Taking the mean of three radii measurements, two from the equator and one from a pole;
	\item Using the radius of a sphere with the same surface area as the Earth's (authalic radius); or
	\item Using the radius of a sphere that has the same volume as the Earth spheroid (volumetric radius).
\end{itemize}

While it is possible to simply use a single value for the radius of the Earth such as a globally averaged value, this will present error in almost any simulation as will be quantified below.  Instead, the value used for $R_{\bigoplus}$ in Equation~\ref{eq:sma} should be relevant to the mean gravitational force that the satellite experiences as it travels around the non-spherical Earth.  In other words, the value for $R_{\bigoplus}$ should be the mean of the distance between the center of the Earth ellipsoid and the surface of the Earth directly under a satellite as it travels in its orbit.  This study will refer to that value as the mean radius under the satellite, or $R_\textrm{s}$. 
Utilizing accurate numbers for this measure is imperative to generating realistic MMs, and consequently realistic TLEs. The following sections detail an approach to determine $R_\textrm{s}$ and provides values that can be used in future simulations.

\section{Mean Radius}
\label{sec:MeanRadius}

This analysis assumes that the Earth is modeled as a WGS-84 ellipsoid\cite{mcclain2001fundamentals}. The Earth ellipsoid also has variation in its gravitational potential as defined in the Earth Gravitational Model 2008\cite{egm2008}
but this study focuses on SGP4 models that include the J2, J3, and J4 zonal harmonics coefficients. This study relies on the use the Satrec package for python\cite{satrec}.  Satrec uses a satellite's TLE to compute the satellite's position and velocity for a given date and time.


To determine $R_\textrm{s}$, a satellite is propagated around the Earth and the geodetic latitude directly under the satellite is determined for many equally-spaced time steps throughout a simulation.  Given each latitude, the radius of the ellipsoid at that position is calculated and stored in an array.  Finally, the mean of these measurements over a simulation yields $R_\textrm{s}$.  In the following sections, we describe a method for determining latitude directly from SGP4 propagation, and then our results for computing $R_\textrm{s}$ from a large set of input inclinations, eccentricities, and arguments of perigee.
\subsection{Determining Radii from Propagation}

Given Cartesian coordinates for a satellite, we can convert them into latitude ($\phi$), longitude ($\lambda$), and height ($h$).  SGP4 propagation provides the ECEF coordinates for a satellite at all desired time steps in the simulation. The longitude is easily computed from the Cartesian coordinates as

\begin{equation}
	\lambda = ArcTan2\left(y,x\right)
\end{equation}

where ArcTan2 is the function that returns the angle from the origin, knowing which quadrant the position $(x,y)$ lies in.  The latitude and height can be computed by application of Ferrari's solution\cite{Zhu} which is not covered here.

For a given latitude $\phi$, the radius of the ellipsoid at that latitude $R_{\phi}$ is given by

\begin{equation}
	R_{\phi} = \sqrt{\frac{\left(R_\textrm{e}^2 cos(\phi)\right)^2 + \left(R_\textrm{p}^2 sin(\phi)\right)^2}{\left(R_\textrm{e} cos(\phi)\right)^2 + \left(R_\textrm{p} sin(\phi)\right)^2}}
\end{equation}

For every time step in the simulation, we determine $\phi$ and therefore a value for $R_{\phi}$. We then take the mean of all values for $R_{\phi}$ to determine $R_\textrm{s}$.

\subsection{Results for $R_\textrm{s}$}
The results in this paper are presented for orbits with inclinations varying from 0 to {90\textdegree}, eccentricities ranging from 0 to 0.999, and arguments of perigee ranging from 0 to {90\textdegree}.  The choice of the altitude at perigee does not affect the results as long it creates a stable orbit.  This analysis considered an altitude at perigee of 605.736~km for all runs. 

For each choice of inclination, eccentricity, and argument of perigee, we determine $R_\textrm{s}$ from a single orbit that is propagated with 1,000 equally spaced time steps. The number of orbits and steps per orbit can be increased but do not have a significant effect on the results.  Furthermore, significantly increasing the number of orbits can introduce unwanted error due to the potentially degraded orbit modeled in SGP4 propagation.  

\begin{figure}[tbph]
	\caption{Top: The mean radius of the Earth, under a satellite, as it propagates throughout its orbit.  The red contour represents the commonly used mean radius of the Earth of 6,371.009\,km.  Bottom: The total percent error between a standard WGS-84 mean Earth radius of 6371.009\,km and the method described in Section~\ref{sec:MeanRadius}. The red contour represents where the difference is exactly zero.  These contour plots were created with an argument of perigee of {0\textdegree}. }
	\includegraphics[width=\linewidth]{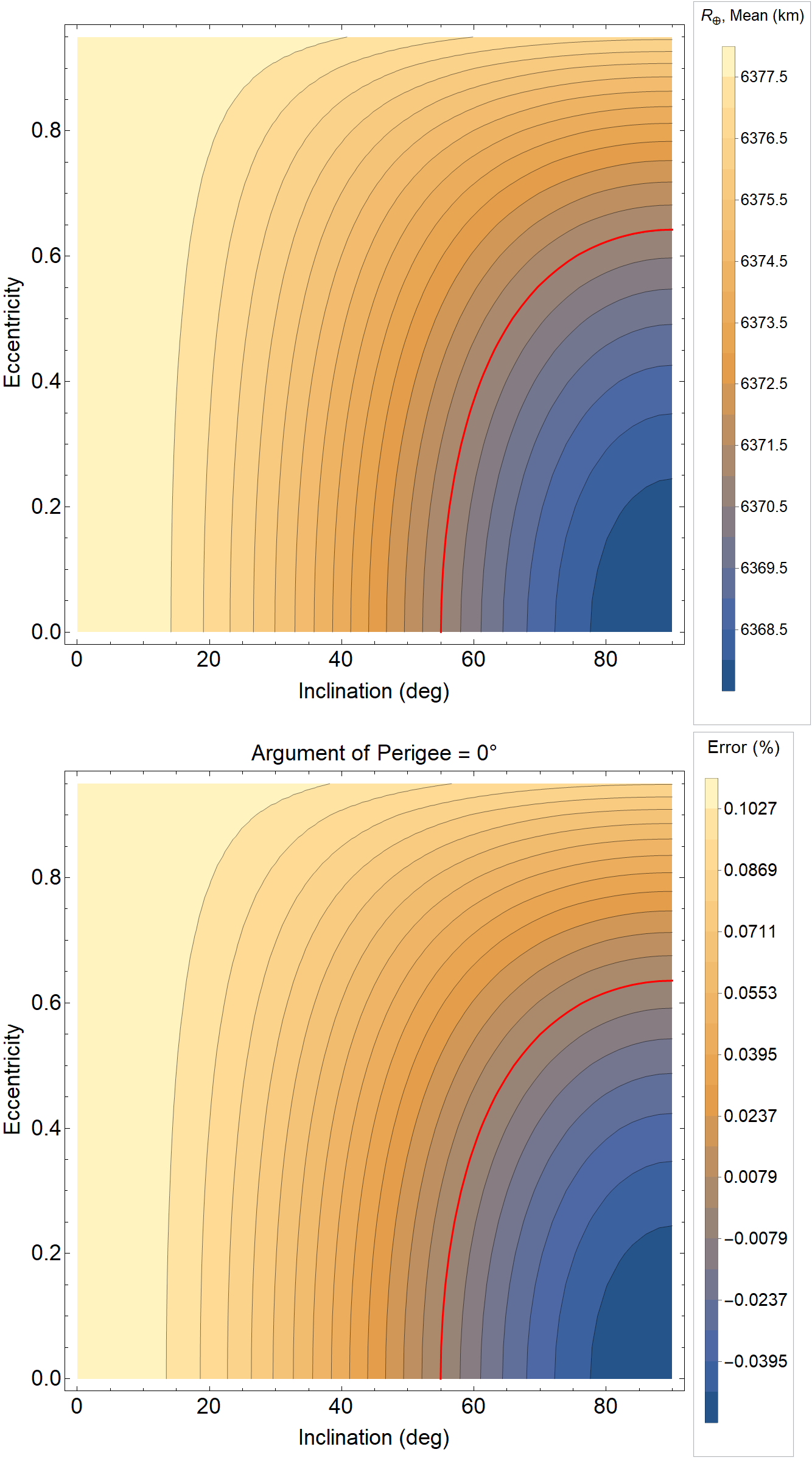}
	\label{im_CP_AP0}
\end{figure}

\begin{figure}[!tb]
	\caption{The time for two TLEs to drift 100 km, each with different assumptions for the Earth radius: one with a standard WGS-84 mean Earth radius, and one with an $R_\textrm{s}$ value computed from Section~\ref{sec:MeanRadius}.  
	This contour plot was created with an argument of perigee of {0\textdegree}.}
	\includegraphics[width=\linewidth]{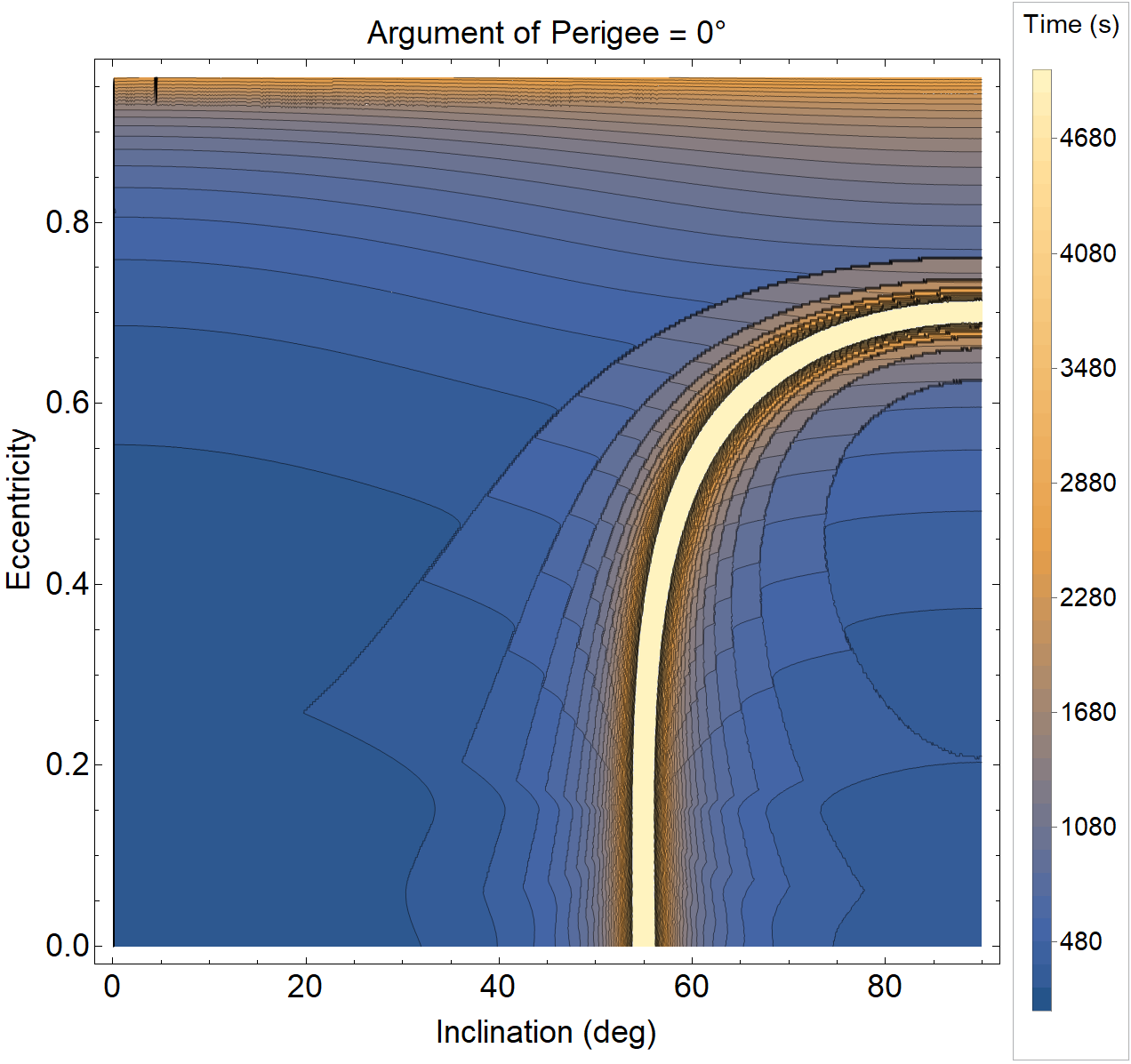}
	\label{im_TimeTo100km}
\end{figure}

The results provide a variation in $R_\textrm{s}$ from 6358.669\,km to 6378.137\,km, partly illustrated in Figure~\ref{im_CP_AP0}. When computing MM, using $R_\textrm{s}$ presents up to a ${\sim}1.5\%$ correction as compared to using the typical WGS-84 mean radius of 6,371\,km.
To measure the impact of such a correction, TLEs are created using both the WGS-84 mean radius of the Earth and $R_\textrm{s}$. This results in two TLEs that are identical except for the MM.  We repeat this process for the same ranges of inclinations, eccentricities, and arguments of perigee.  The time for each pair of TLEs to drift 100 km apart is measured and illustrated in Figure~\ref{im_TimeTo100km}.  

In many cases, this drift can be greater than 100 km in less than only a few minutes. For many analyses such as conjunction analysis, object tracking, and coverage metrics, this accumulated error can present significant changes to the simulation results, especially for longer duration simulations. The method described above for determining $R_\textrm{s}$ presents an improved means of positioning and modeling satellites through longer simulations. 

\section{Assembling Terms for Notional TLEs}
\label{sec:assemble}
With the catalog-derived terms ready, realistic TLEs can be created based on the desired inclination, eccentricity, argument of perigee, RAAN, and altitude at perigee.  Table~\ref{table:TLETerms} lists all TLE terms, and whether they are derived from a catalog, determined by calculation, or taken as direct inputs.
\begin{table}
	\centering
	\begin{tabular}{|l|c|}
		\hline
		\Tstrut Term & Method\\\hline
		First Derivative of MM & Catalog\\
		Second Derivative of MM & Catalog\\
		Drag (B*) & Catalog \\ 
		Inclination & Free \\
		RAAN & Free \\
		Eccentricity & Free \\
		Argument of Perigee & Free \\
		Mean Anomaly & Free\\
		Mean Motion & Determined from $R_\textrm{s}$\Bstrut\\\hline
	\end{tabular}
	\caption{The TLE terms used in this study, and whether they are catalog-derived, free variables, or determined from other elements including the calculation of $R_\textrm{s}$ as described in Section~\ref{sec:MeanRadius}.  All other TLE terms are trivially chosen except for the checksum which is computed based on all other entries in each line.\label{table:TLETerms}}
\end{table}

The procedure for creating a TLE is as follows:
\begin{enumerate}
	\item The desired inclination, eccentricity, altitude at perigee, and argument of perigee are chosen according to the desired orbit
	\item The $R_\textrm{s}$ value is determined based inclination, eccentricity, and argument of perigee
	\item The semi-major axis is determined from Equation~\ref{eq:sma}, substituting $R_\textrm{s}$ in place of $R_{\bigoplus}$
	\item The MM is calculated from Equation~\ref{eq:MM}
	\item The satellite population is classified according to Algorithm~\ref{alg:TLEClassification} using the calculated MM and intended eccentricity
	\item The first and second derivative of mean motion and drag terms are assigned from Table~\ref{table:1st2ndBstar}
	\item The rest of the TLE elements are freely chosen, and the checksum is computed for the last character in each line
\end{enumerate}

The simulated data for $R_\textrm{s}$ described in Section~\ref{sec:MeanRadius} can be fit by a polynomial function whose parameters can be numerically determined.  We fit a fifth-order 3-dimensional polynomial to the data for $R_s$ with 56 parameters and determined a maximum difference of $5.47\times10^{-3}$\,\% compared to directly simulating the orbit.  This is across all inclinations, eccentricities, and arguments of perigee.  Similarly, for an eighth-order polynomial with 165 parameters, the maximum difference was $0.93\times10^{-3}$\,\% compared to direct simulation, across all values. The values for the parameters of the fifth-order polynomial are shown in Table~\ref{table:5thOrder}.  The polynomial with associated parameters is shown in Equation~\ref{eq:fifthOrder} with $i$, $e$, and $\omega$ the inclination, eccentricity, and argument of perigee, respectively.

\begin{equation}
	\begin{split}
		\textrm{R}_\textrm{s}\left(i,e,\omega\right)=&c_{000}+c_{001}\omega+c_{002}\omega^2+c_{003}\omega^3+c_{004}\omega^4+\\
		c_{005}\omega^5+&c_{010}e+c_{011}{\omega}e+c_{012}{\omega}^2e+c_{013}{\omega}^3e+\\
		c_{014}{\omega}^4e+&c_{020}e^2+c_{021}{\omega}e^2+c_{022}{\omega}^2e^2+c_{023}{\omega}^3e^2+\\
		c_{030}e^3+&c_{031}{\omega}e^3+c_{032}{\omega}^2e^3+c_{040}e^4+c_{041}{\omega}e^4+\\
		c_{050}e^5+&c_{100}i+c_{101}i{\omega}+c_{102}i{\omega}^2+c_{103}i{\omega}^3+\\
		c_{104}i{\omega}^4+&c_{110}ie+c_{111}ie{\omega}+c_{112}ie{\omega}^2+c_{113}ie{\omega}^3+\\
		c_{120}ie^2+&c_{121}ie^2{\omega}+c_{122}ie^2{\omega}^2+c_{130}ie^3+\\
		c_{131}ie^3{\omega}+&c_{140}ie^4+c_{200}i^2+c_{201}i^2{\omega}+c_{202}i^2{\omega}^2+\\
		c_{203}i^2{\omega}^3+&c_{210}i^2e+c_{211}i^2e{\omega}+c_{212}i^2e{\omega}^2+\\
		c_{220}i^2e^2+&c_{221}i^2e^2{\omega}+c_{230}i^2e^3+c_{300}i^3+c_{301}i^3{\omega}+\\
		c_{302}i^3{\omega}^2+&c_{310}i^3e+c_{311}i^3e{\omega}+c_{320}i^3e^2+c_{400}i^4+\\
		c_{401}i^4{\omega}+&c_{410}i^4e+c_{500}i^5
		\label{eq:fifthOrder}
	\end{split}
\end{equation}

\begin{table}[!tb]
	\centering
	\small
	\begin{tabular}{|l|c||l|c|}
		\hline
		\Tstrut Term & Value & Term & Value\\\hline\Tstrut
		$c_{000}$ & $6.377788\times10^6$ & $c_{112}$ & $-9.972427\times10^{-2}$\\
		$c_{001}$ & $2.836106\times10^1$ & $c_{113}$ & $7.383957\times10^{-4}$\\
		$c_{002}$ & $-4.915346\times10^{-1}$ & $c_{120}$ & $9.150674\times10^{1}$\\
		$c_{003}$ & $-4.017214\times10^{-3}$ & $c_{121}$ & $-1.33747\times10^{0}$\\
		$c_{004}$ & $1.273759\times10^{-4}$ & $c_{122}$ & $1.155062\times10^{-3}$\\
		$c_{005}$ & $-5.643483\times10^{-7}$ & $c_{130}$ & $4.645793\times10^{0}$\\
		$c_{010}$ & $2.120874\times10^3$ & $c_{131}$ & $-1.909859\times10^{0}$\\
		$c_{011}$ & $-1.886565\times10^2$ & $c_{140}$ & $5.290634\times10^{1}$\\
		$c_{012}$ & $5.241726\times10^0$ & $c_{200}$ & $-3.926748\times10^{0}$\\
		$c_{013}$ & $-3.84173\times10^{-2}$ & $c_{201}$ & $1.703196\times10^{-2}$\\
		$c_{014}$ & $-2.268633\times10^{-6}$ & $c_{202}$ & $-8.207037\times10^{-6}$\\
		$c_{020}$ & $-3.447518\times10^3$ & $c_{203}$ & $2.674298\times10^{-8}$\\
		$c_{021}$ & $8.553227\times10^1$ & $c_{210}$ & $4.368207\times10^{0}$\\
		$c_{022}$ & $-4.719285\times10^0$ & $c_{211}$ & $-9.744218\times10^{-2}$\\
		$c_{023}$ & $3.438466\times10^{-2}$ & $c_{212}$ & $-8.298287\times10^{-6}$\\
		$c_{030}$ & $9.952275\times10^3$ & $c_{220}$ & $2.116514\times10^{-1}$\\
		$c_{031}$ & $1.634122\times10^2$ & $c_{221}$ & $-4.547244\times10^{-4}$\\
		$c_{032}$ & $5.259614\times10^{-2}$ & $c_{230}$ & $-9.768337\times10^{-2}$\\
		$c_{040}$ & $-1.695787\times10^4$ & $c_{300}$ & $1.323687\times10^{-3}$\\
		$c_{041}$ & $-8.008678\times10^1$ & $c_{301}$ & $-1.327782\times10^{-4}$\\
		$c_{050}$ & $8.501297\times10^3$ & $c_{302}$ & $5.032226\times10^{-8}$\\
		$c_{100}$ & $2.948006\times10^1$ & $c_{310}$ & $-3.319618\times10^{-2}$\\
		$c_{101}$ & $-1.151884\times10^0$ & $c_{311}$ & $7.264131\times10^{-4}$\\
		$c_{102}$ & $1.653916\times10^{-2}$ & $c_{320}$ & $-5.846808\times10^{-4}$\\
		$c_{103}$ & $-1.276845\times10^{-4}$ & $c_{400}$ & $4.655915\times10^{-4}$\\
		$c_{104}$ & $3.521573\times10^-8$ & $c_{401}$ & $6.438237\times10^{-8}$\\
		$c_{110}$ & $-1.676297\times10^2$ & $c_{410}$ & $4.999741\times10^{-6}$\\
		$c_{111}$ & $6.645437\times10^0$ & $c_{500}$ & $-2.08623\times10^{-6}$\Bstrut\\\hline
	\end{tabular}
	\caption{The parameters for a fifth-order polynomial fit of the $R_\textrm{s}$ data across all inclinations, eccentricities, and arguments of perigee from Section~\ref{sec:MeanRadius}.\label{table:5thOrder}}
\end{table}

\section{Conclusion}

We present a new means of creating TLEs for notional satellites with realistic terms for mean motion, its first and second derivatives, and the drag term. The only inputs required are the desired inclination, eccentricity, argument of perigee, and altitude at perigee.  The produced TLEs can be used in simulation alone or alongside TLEs from real satellites, and will behave like real systems.  This increases the overall simulation accuracy for performance evaluation, conjunction analysis, and more.  The methods described can be expanded to other classification schemes for different sets of populations.

Future work will involve investigating the satellite catalog and possibly deriving correlations among populations that could reduce the standard deviations in Table~\ref{table:1st2ndBstar}, though this may mean significantly increasing the complexity of population categorization. This work could also be adapted to include the EGM2008 terms and also regenerated for different satellite classification schemes\cite{egm2008}. 

\section*{\textit{Acknowledgments}}
This technical data was produced for the U. S. Government under Contract No. FA8702-22-C-0001, and is subject to the Rights in Technical Data-Noncommercial Items Clause DFARS 252.227-7013 (FEB 2014). \textcopyright~2022 The MITRE Corporation. All Rights Reserved. DISTRIBUTION A: Approved for public release; distribution unlimited. Public Release Case Number 22-0304.

\bibliographystyle{unsrtnat}
\bibliography{ReferencesTLEMaker}

\begin{thebibliography}{13}
\providecommand{\natexlab}[1]{#1}
\providecommand{\url}[1]{\texttt{#1}}
\expandafter\ifx\csname urlstyle\endcsname\relax
  \providecommand{\doi}[1]{doi: #1}\else
  \providecommand{\doi}{doi: \begingroup \urlstyle{rm}\Url}\fi

\bibitem[UCS(2021)]{UCSSatelliteDatabase_2021}
Ucs satellite database, May 2021.
\newblock URL \url{https://www.ucsusa.org/resources/satellite-database}.
\newblock \url{https://www.ucsusa.org/resources/satellite-database}.

\bibitem[Parish(2004)]{Coverage1}
Jason~A. Parish.
\newblock Optimizing coverage and revisit time in sparse military satellite
  constellations; a comparison of traditional approaches and genetic
  algorithms, 2004.
\newblock URL \url{https://calhoun.nps.edu/handle/10945/1209}.

\bibitem[Hitomi and Selva(2018)]{8396743}
Nozomi Hitomi and Daniel Selva.
\newblock Constellation optimization using an evolutionary algorithm with a
  variable-length chromosome.
\newblock In \emph{2018 IEEE Aerospace Conference}, pages 1--12, 2018.
\newblock \doi{10.1109/AERO.2018.8396743}.

\bibitem[Budianto and Olds(2004)]{Design}
Irene~A. Budianto and John~R. Olds.
\newblock Design and deployment of a satellite constellation using
  collaborative optimization.
\newblock \emph{Journal of Spacecraft and Rockets}, 41\penalty0 (6):\penalty0
  956--963, 2004.
\newblock \doi{10.2514/1.14254}.
\newblock URL \url{https://doi.org/10.2514/1.14254}.

\bibitem[Vallado et~al.(2012)Vallado, Crawford, Hujsak, and Kelso]{SGP4}
David Vallado, Paul Crawford, Richard Hujsak, and T.S. Kelso.
\newblock Revisiting spacetrack report \#3.
\newblock \emph{AIAA/AAS Astrodynamics Specialist Conference and Exhibit},
  2012.
\newblock \doi{10.2514/6.2006-6753}.
\newblock URL \url{https://arc.aiaa.org/doi/abs/10.2514/6.2006-6753}.

\bibitem[Kelso(2019)]{kelso_2019}
T~S Kelso.
\newblock Norad two-line element set format, Dec 2019.
\newblock URL \url{https://www.celestrak.com/NORAD/documentation/tle-fmt.php}.
\newblock \url{https://www.celestrak.com/NORAD/documentation/tle-fmt.php}.

\bibitem[Vallado and Cefola(2012)]{ValladoTLE}
David Vallado and Paul Cefola.
\newblock Two-line element sets - practice and use.
\newblock \emph{Proceedings of the International Astronautical Congress, IAC},
  7:\penalty0 5812--5825, 01 2012.

\bibitem[Mamajek et~al.(2015)Mamajek, Prsa, Torres, Harmanec, Asplund, Bennett,
  Capitaine, Christensen-Dalsgaard, Depagne, Folkner, et~al.]{Mamajek2015}
EE~Mamajek, A~Prsa, G~Torres, P~Harmanec, M~Asplund, PD~Bennett, N~Capitaine,
  J~Christensen-Dalsgaard, E~Depagne, WM~Folkner, et~al.
\newblock Iau 2015 resolution b3 on recommended nominal conversion constants
  for selected solar and planetary properties, arxiv151007674 astro-ph, 2015.

\bibitem[Moritz(1980)]{Moritz1980}
Helmut Moritz.
\newblock Geodetic reference system 1980.
\newblock \emph{Bulletin g{\'e}od{\'e}sique}, 54\penalty0 (3):\penalty0
  395--405, 1980.

\bibitem[McClain and Vallado(2001)]{mcclain2001fundamentals}
W.D. McClain and D.A. Vallado.
\newblock \emph{Fundamentals of Astrodynamics and Applications}.
\newblock Space Technology Library. Springer Netherlands, 2001.
\newblock ISBN 9780792369035.
\newblock URL \url{https://books.google.com/books?id=PJLlWzMBKjkC}.

\bibitem[Pavlis et~al.(2012)Pavlis, Holmes, Kenyon, and Factor]{egm2008}
Nikolaos~K. Pavlis, Simon~A. Holmes, Steve~C. Kenyon, and John~K. Factor.
\newblock The development and evaluation of the earth gravitational model 2008
  (egm2008).
\newblock \emph{Journal of Geophysical Research: Solid Earth}, 117\penalty0
  (B4), 2012.
\newblock \doi{https://doi.org/10.1029/2011JB008916}.
\newblock URL
  \url{https://agupubs.onlinelibrary.wiley.com/doi/abs/10.1029/2011JB008916}.

\bibitem[Rhodes(2021)]{satrec}
Brandon Rhodes.
\newblock python-sgp4.
\newblock \url{https://github.com/brandon-rhodes/python-sgp4}, 2021.

\bibitem[Zhu(1994)]{Zhu}
J.~Zhu.
\newblock Conversion of earth-centered earth-fixed coordinates to geodetic
  coordinates.
\newblock \emph{IEEE Transactions on Aerospace and Electronic Systems},
  30\penalty0 (3):\penalty0 957--961, 1994.
\newblock \doi{10.1109/7.303772}.

\end{thebibliography}

\end{document}